# Pervasive Flexibility in Living Technologies through Degeneracy Based Design

*James Whitacre\**, Adaptive Risk Management Lab, School of Computer Science, University of New Brunswick, Fredericton, Canada. Email: jwhitacre79@gmail.com

*Axel Bender*, Land Operations Division, Defence Science and Technology Organisation, Edinburgh, SA, Australia, and School of Engineering and Information Technology, University of New South Wales, Canberra, ACT, Australia. axel.bender@dsto.defence.gov.au

*corresponding author

**Abstract:** The capacity to adapt can greatly influence the success of systems that need to compensate for damaged parts, learn how to achieve robust performance in new environments, or exploit novel opportunities that originate from new technological interfaces or emerging markets. Many of the conditions in which technology is required to adapt cannot be anticipated during its design stage, creating a significant challenge for the designer. Inspired by the study of a range of biological systems, we propose that degeneracy – the realization of multiple, functionally versatile components with contextually overlapping functional redundancy – will support adaptation in technologies because it effects pervasive flexibility, evolutionary innovation, and homeostatic robustness. We provide examples of degeneracy in a number of rudimentary living technologies from military socio-technical systems to swarm robotics and we present design principles – including protocols, loose regulatory coupling, and functional versatility – that allow degeneracy to arise in both biological and man-made systems.

# 1. Introduction

Unanticipated requirements can arise throughout a technology's life and are a notoriously difficult engineering problem and a challenging research topic because past routines and contingency plans will be of limited utility. Dealing with new challenges requires exploration, diversity, and bet-hedging: principles that are common to any discipline in which responses to novelty determine competitive success.

However these conceptualizations of adaptive behaviour provide only approximate descriptive accounts of how complex technologies can be designed to achieve robustness and adaptation in novel circumstances. Importantly, it is still poorly understood how adaptive options are generated and exploited within a systems context without sacrificing other objectives related to efficiency and effectiveness. Recent theories have proposed that a biological property known as degeneracy plays an important role in establishing synergistic relationships between biosystem complexity, adaptive potential, robustness, and efficiency [1]. In this paper, we will explore how the development of degeneracy in living technologies [2] may help these systems acquire desirable adaptive features. We will propose a precise form of diversity using degeneracy based design principles that can support distributed robustness in a number of social and technological multi-agent systems. While not a complete recipe for realizing the living technology vision, the principles emphasized here have been chosen because they are ubiquitous in evolvable biosystems, are mostly absent in fragile and non-adaptive designed systems, and are notably absent in all but a few discussions that address the requisite conditions of artificial life [3], artificial intelligence [4], and living technologies [5, 6].

# 2. The Design and Development of Flexible Systems

*If flexibility only exists in places where we perceive future need, then our resilience will be limited by our foresight. Because novel requirements are not predictable, flexible responses to novelty cannot be entirely pre-specified. Instead, flexibility must be a pervasive property that emerges on demand without explicit planning or foresight.*

Researchers and engineers from largely isolated disciplines have uncovered similar principles that contribute to the design or evolution of adaptive systems and appear to be widely applicable within ecosystems, biochemical networks, systems engineering, and human organizations [7-10]. Primary factors that contribute to resilience of food-webs, canalization of multi-cellular development, physiological homeostasis, and robust control of automated manufacturing processes include intuitive engineering concepts such as functional redundancy, bet-hedging, saturation effects, and fail-safe principles.

Feedback control concepts have been particularly successful in explaining robustness in a wide range of systems including biological networks [11-14]. These findings have received considerable publicity in biology, in part because they support the view of a top-down deconstruction of biological complexity that could reveal intuitive insights into the exceptional robustness and adaptive potential of living systems [15].

Less attention has been given to evidence that the majority of robustness in biological systems cannot be attributed to simple feedback loops or to perfect redundancy [9, 10, 16-21]. Even though control theoretic principles are relevant to biological robustness, the sets of components displaying single reference point, closed-loop control principles only do so within tightly restricted (microenvironment) conditions and it is more often the case that numerous interdependent actions amongst diverse biological elements are integrated over space and time with robustness emerging in a distributed fashion. In biology, this phenomenon is referred to as distributed robustness [18, 19] or emergent flexibility [5].

Biological and ecological research on distributed robustness has uncovered statistical patterns of regulatory (activation/inhibition) and mass-action interactions that are positively correlated with robustness including nested feedback loops, bow-tie architectures, and long-tail distributions of regulatory interactions [16, 22, 23]. For instance, in the immune system [24] and metabolism [22], distributed robustness is facilitated in part by a multi-scaled bow-tie architecture: at many scales of the system there exist multiple pathways to achieving a given function/effect. The result of these multiple pathways is that the system is endowed with exceptional flexibility when operating under stressed conditions. Interestingly, these pathways are compensatory but not entirely redundant: in many circumstances they contribute to entirely different functions. As a simple example, the metabolism of glucose can take place through two distinct pathways; glycolysis and the pentose phosphate pathway. Although these pathways can substitute for each other when necessary, the entirety of their metabolic effects is not identical. Distributed robustness can emerge in similar ways within human organizational contexts. For instance, military adaptive capabilities arise within networked force elements that compensate and complement each other. This allows for a changeable organizational form that emerges in response to deployment contexts of large (and dangerous) uncertainty.

Within socio-technical systems, it is not controversial to assert that adequate responses toward novel internal and external stresses generally require flexibility in what/when/where actions are taken by a combination of human, hardware, and electronic assets. However, because the what/when/where of novel requirements is fundamentally unpredictable, the flexibility needed to respond to this novelty cannot be pre-specified based on the anticipation of future conditions. Instead, the adequate provision of flexibility requires it to be a pervasive system property that can emerge without explicit planning or foresight. Importantly, if flexibility only arises in the places where we perceive future need then resilience will be limited by our foresight, e.g. our ability to predict plausible future scenarios.

A rich history of engineering and planning experience suggests that pervasive flexibility is prohibitively costly and impractical due to the inefficiency of idle redundant resources (robustness-efficiency trade-off). The alternative, where resource flexibility emerges on demand without requiring idle/backup resources, is not easily imagined within a technological or engineering context. In some circumstances, flexibility requires diversity in options, not redundancy, however there are management overhead costs from diversity that add to system complexity and should slow-down the speed of adaptation in large systems (complexity-adaptation trade-off). Drawing from these perspectives, pervasive flexibility is assumed to only be possible in nature because natural selection permits inefficiency and a slow pace of adaptation. The problem with these assumptions is that they are inaccurate and misleading: biological systems evolve within highly competitive and resource-constrained environments and rapid evolutionary change is common in even the most complex species [25]. Biological systems bypass or partly resolve the conflicts between robustness, efficiency, complexity, and evolvability that limit technological capabilities [1].

Although survival and fecundity are not perfect analogues to market-based forces, some researchers believe that the similarities are sufficient in some circumstances to warrant research into the nature-inspired design of artificial systems. Recently we proposed a theory to explain how pervasive forms of biological flexibility are achieved at high levels of efficiency through a property known as degeneracy [19]. Degeneracy refers to a unique situation where groups of agents will exhibit functional redundancy in some contexts but functional diversity in others.

## 3. Contributing Factors for Adaptation in Biological, Social, and Engineered Systems

To understand how degeneracy can achieve the proposed effects, it is helpful to first discuss the typical options that are available to a system that is responding to novel conditions. As listed in Figure 1 and Table 1, response options typically involve one or a combination of the following:

1. reducing exposure to unwanted conditions by manipulating the environment;
2. reducing exposure to unwanted conditions through mobility within environment;
3. adapting system behaviour in response to environmental change in order to improve, maintain, or repair a capability output;
4. designing high quality components that are individually robust towards stress and damage. In this case, no adaptive response is required.

Each of these options may invoke prototypical examples in the reader's mind, however each is widely applicable across many systems contexts. Response options are a prerequisite for adaptation and involve one or a combination of the following actions:

1. **Changes in how much, when, and where resources are needed**: This form of adaptation requires options to quickly change the quantity of a particular functional output at a particular place and time. Excess backup resources can support this type of adaptation, however idle resources reduce average efficiency and thus can be costly.
2. **Changes in task specifications**: unexpected conditions sometimes require a function to be executed in a manner that deviates from the norm. Maintaining diversity in the options for executing a task, with each option displaying unique vulnerabilities, can provide reliability in the face of novel requirements. Option diversity is typically not random and instead reflects an accumulated knowledge of expected disturbances. For instance, bet-hedging strategies drawn from portfolio theory are used in several disciplines to reduce the likelihood of large systemic risks against known uncertainties [26-28].
3. **Functional novelty (exaptation)**: New environments can reveal opportunities to utilize existing components in novel ways: a class of adaptation that is known to biologists as exaptation [29]. Maintaining diversity in options/assets/agents can improve the likelihood of discovering exaptation opportunities. Successful R&D departments (e.g. at 3M, Apple, Google) are often in the habit of exploring potential exaptations for existing products or organizational competencies.

In summary, adaptation involves several distinct requirements, in some cases demanding quantitative changes in functional outputs through redundancy while in other cases demanding qualitatively diverse options to sustain functional capabilities in novel environments or to explore new capability opportunities. In systems engineering and organization science, these requirements for adaptation have been addressed through largely separate programs involving redundancy, bet-hedging, exploration, and related strategies in Figure 1 and Table 1.

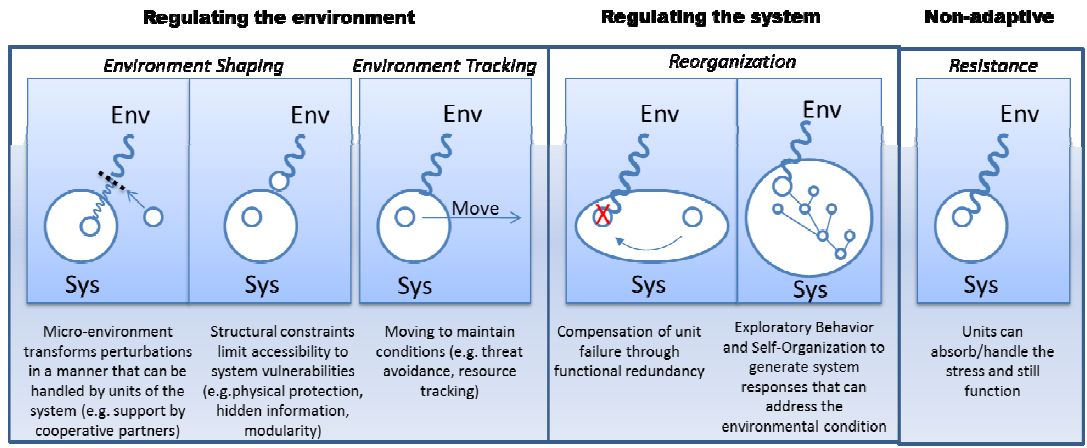

Figure 1: four pathways for adapting to environmental stress: environmental shaping – regulate the environment so fragilities are not accessed/revealed; environmental tracking – move to new environments where performance can be maintained; reorganization – adapt system response in order to sustain or improve upon system objectives; resistance – design high quality components that resist damage from perturbations

Table 1: factors that contribute to robust system performance under volatile conditions

| Mechanisms and properties that enhance robustness | Biological Examples | Engineering and Management Science Examples | Military Examples |
| --- | --- | --- | --- |
| **Reliability through functional and pathway redundancy** (distinct components/pathways that are interchangeable and thus robust against the loss of a single component) | Gene regulation, protein functionality, metabolic and signalling pathways, and neural anatomy can be highly degenerate and thus display some degree of functionally redundancy. | Empirically driven placement of backup devices as well as storage/maintenance/preservation facilities can buffer against fluctuating operating conditions or component failure. Feedback diagnostics for preventative maintenance to replace components before they fail. | Backup communications, excess resources, and multiple options for completing a mission all provide reliability under uncertain conditions |
| **Resistance** (robustness of component towards variable conditions removes need for any system level response) | Many types of threshold effects in biology appear as sub-systems with innate (albiet bounded) resistance to change (e.g. Genetic switches, TCR mediated activation of T cells, neural activation) | High cost ultraquality components with lower rates of failure can provide reliability in circumstances where replacement is impractical. | Rugged high performance equipment is a common feature of defence hardware |
| **Local environment shaping /regulation** (Instead of achieving robustness by responding to environmental stress, it is sometimes possible to shape the environment in ways that allow a system to avoid exposure to damaging stress) | Niche construction and environment simplification alter the type and frequency of perturbations encountered. Heat shock proteins (e.g. Hsp90) assist other proteins to fold and refold into functionally relevant conformations and confer conformational robustness toward thermal fluctuations and canalize a broad range of morphological traits {Rutherford, 1998 #966}. Localization of harmful pathogens through tissue inflamation or through ingestion by macrophages | Monitoring and controlling sub-system operating environments can reduce exposure to damaging perturbations. Fail-safe principles can dynamically encapsulate subsystems (i.e. dynamically constructed modularity) and prevent failures from propagating into expensive devices and system critical operations. | Armour, secure/safe zones, bunkers, provide protection to otherwise vulnerable assets. Engagement with local communities helps to shape risks and resources |
| **Mobility** (having the ability to move or be moved into environments can enable functions to be achieved when conditions demand them or to be relocated when hostile conditions develop) | Predator avoidance, adaptive foraging, migration, and seed dispersal all provide options for populations to seek out and track suitable habitats. | | A considerable amount of military hardware has the express purpose of providing mobility in various circumstances |

# 4. Degeneracy

The distinct forms of adaptation discussed in the previous sections are each partly supported in biological systems through degeneracy. Degeneracy is a property seen in repertoires of multi-functional agents when some of the agents are functionally interoperable for certain types of tasks but uniquely functionally qualified for others (Figure 2).

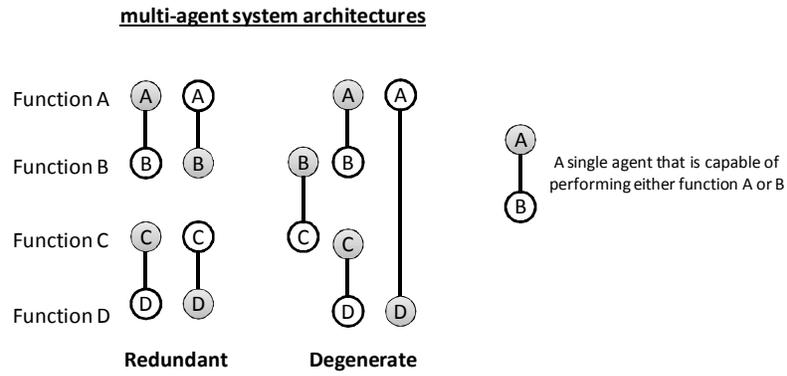

Figure 2: multi-functional agents are shown that are either perfectly identical in functional capabilities (purely redundant) or partially redundant (degenerate).

Degeneracy is not restricted to biological systems and can be easily seen in many complex adaptive systems. Conceptual illustrations of degeneracy are given in Figure 3 for small and large defence systems including military field vehicles, joint operations, and multi-nation alliances. Additional biological and human organizational examples are listed in Table 2.

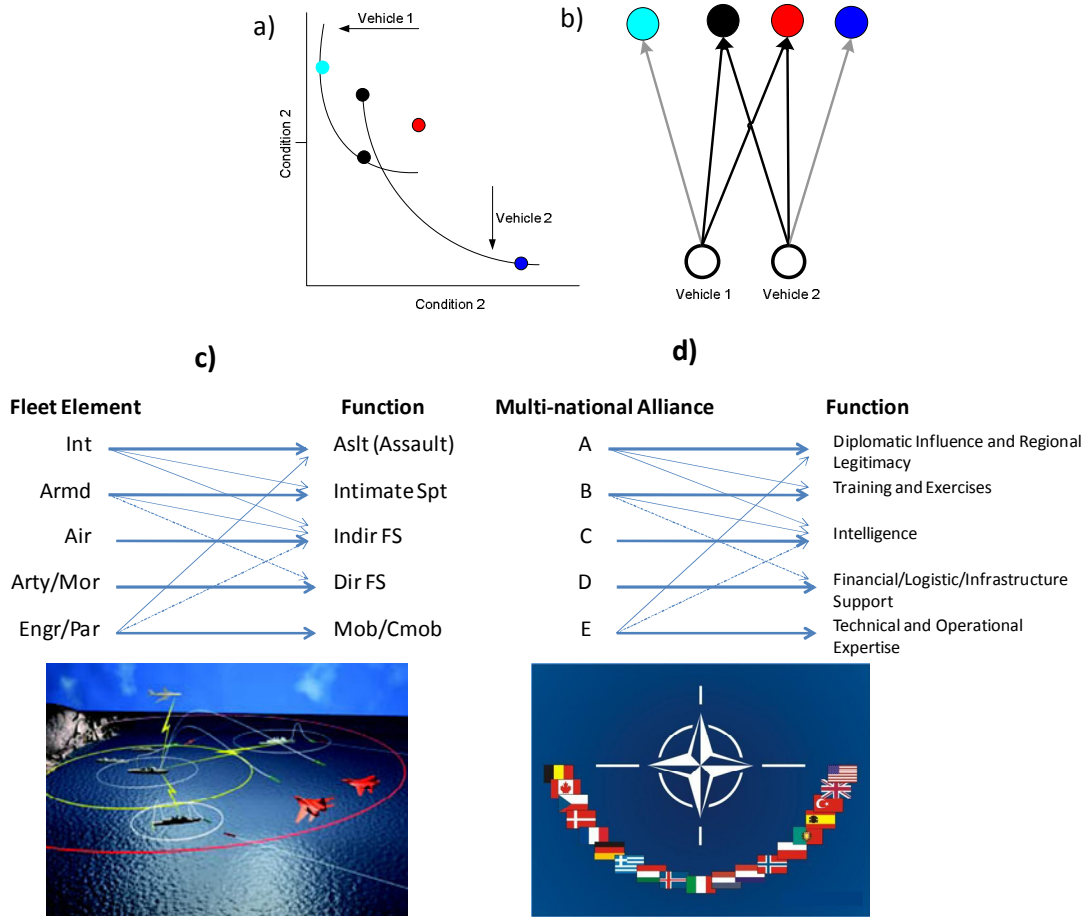

Figure 3 Examples of degeneracy at different levels of organization in Defence. Panel a) Different vehicle types have unique tradeoffs that define the conditions under which they can operate. The continuous tradeoff surface (arcs) conceptually illustrates a vehicle's operational range. The red node typifies a set of conditions under which either vehicle type is suitable. The black nodes represent a task where condition two is a hard constraint and condition one is a soft constraint. Either vehicle can perform the task but at different performance levels. Blue nodes are tasks where each vehicle is uniquely qualified. Panel b) Vehicles 1 and 2 from Panel a are degenerate, i.e. under certain conditions the vehicles are interoperable while in others the vehicles are uniquely qualified. Panel c) Degeneracy in the relationship between defence force elements and combat functions. Panel d) Degeneracy in the relationship between nations and strategic capabilities.

Table 2 System classes where agents are multifunctional and have functions that can partially overlap with other agents. Degeneracy is observed in each case through the conditional interoperability in tasks.

| Agent | System | Environment | Control | Agent Tasks |
|---|---|---|---|---|
| Vehicle type | Transportation Fleet | Transportation Network | Centralized Command and Control | Transporting goods, pax |
| Force element | Defence Force Structure | Future Scenarios | Strategic Planning | Missions |
| Person | Organization | Marketplace | Management | Job Roles |
| Deme | Ecosystem | Physical Environment | Self-organized | Resource usage and creation |
| Gene Product | Interactome | Cell | Self-organized and evolved | Energetic and sterric interactions |
| Antigen | Immune System | Antibodies and host proteins | Immune learning | Recognizing foreign proteins |

# 5. The logical and intuitive benefits from degeneracy

Some of the desirable properties that arise from degeneracy can be related to simple concepts such as functional redundancy, bet-hedging, and exploratory behaviour. One simple way degenerate components contribute to adaptation is through component multi-functionality. By being able to contribute to a variety of tasks, multi-functional components can change what they do and contribute to system responses involving quantitative changes in functional outputs. Simply stated, multi-functional components can engage in one particular function if more resources for that function are needed or be reassigned to one of its other functions if fewer of those resources are needed. Thus, multi-functional agents can support adaptive responses to changing task requirements (Figure 4a,b).

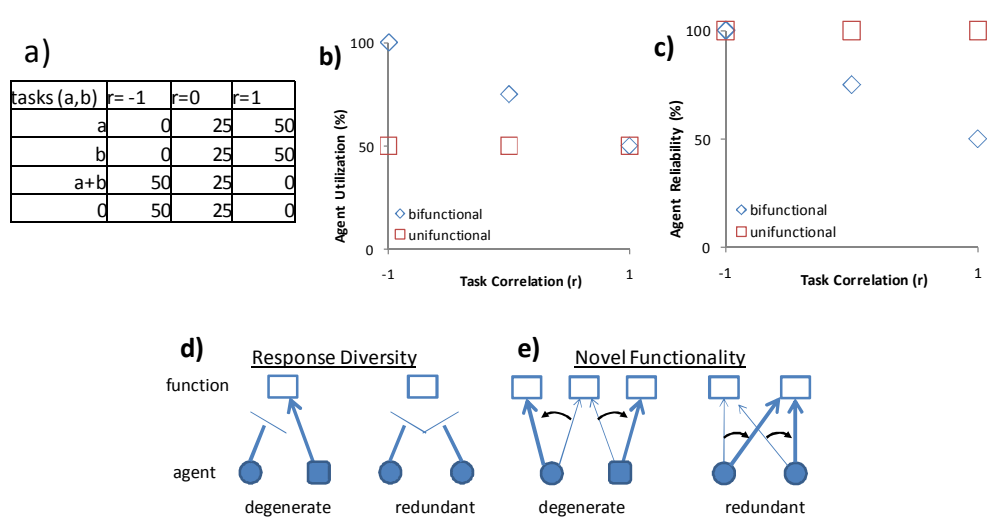

**Figure 4 Panel a,b) Multi-functional components are less likely to sit idle under fluctuating task requirements. Panel a)** probability of events where each task (a,b) has a 50% probability of occurring and tasks are 100% positively correlated (r=1), uncorrelated (r-0), and 100%negatively correlated (r=-1). **Panel b)** Expected utilization rate for bi-functional (a+b) and uni-functional (a or b) agents. **Panel c)** Expected availability for a specific task. **Panel d)** structural differences can enhance reliability in executing a function. **Panel e)** A greater variety of novel functions can be revealed when degenerate agents are placed in new environments.

Components that are degenerate can also support bet-hedging and exploratory adaptive responses. First, degenerate components are functionally interchangeable for certain tasks but achieve these tasks in different ways. These differences can lead to differences in agent performance when tasks are executed under novel or rare conditions. Because of this context-revealed "response diversity" (Figure 4d), satisfying a particular task under unexpected novel conditions is more likely to be possible by a repertoire of degenerate versus redundant components. In a similar vein, degenerate components can harbour somewhat distinct vulnerabilities thus increasing the likelihood that at least one component will not fail when confronted with a novel or rare disturbance, thus providing a basic form of bet-hedging.

Novel environments sometimes reveal opportunities for a component to be co-opted to perform a new beneficial function. With behavioral differences amongst degenerate components, each component harbours a contextually unique potential for co-option. A group of degenerate elements thus provides greater opportunities for exploring innovative capabilities and for responding to novel functional requirements.

## Emergent benefits from degeneracy

Degeneracy can also support emergent forms of distributed flexibility that could be relevant to the adaptive capabilities of living technologies. First, when multi-functional components are interoperable in only a subset of their functions (i.e. degenerate), fluctuating task requirements can cause interoperability options to become synergistically linked and result in a basic form of distributed robustness. An example of this "synergistic linkage" is illustrated in Figure 5a. In the figure, Agent B can perform either task 2 or 3. If Agent B has no task assigned to it, then it is available to take over tasks (of type 2 or 3) assigned to Agents A or C. This allows Agents A and C to be available for tasks (of type 1 and 4) that Agent B could not carry out. In other words, resources of Agent B not only support adaptation toward variable demands in tasks 2 and 3, they can also indirectly enable new resources to be available for tasks 1 and 4: tasks that are unrelated to Agent B. While perfect interoperability deconstrains which agents are assigned to a given task, partial interoperability allows excess resources related to a particular task to support fluctuations in unrelated tasks.  As a result, small amounts of local excess resources can be utilized in a highly versatile manner, thereby increasing the variety of task fluctuations that a system can respond to. We have shown that this simple effect can fundamentally alter the trade-off between robustness and efficiency [19].

In particular, we have discovered that some architectures allow systems to respond to enormous task fluctuations; a phenomena that we described as the Networked Buffering (NB) hypothesis [19]. According to NB, if partial interoperability relationships form a connected network (Figure 5c), then the repertoire of response options towards different task requirements can become very large. This is qualitatively illustrated in Figure 5c and quantitatively validated in multi-agent simulations in Figure 5d. In contrast, when multi-functional components are designed/trained so that functions are clustered within organizational stovepipes, a multiplier effect is not observed (Figure 5b,d).  The distributed robustness just described is arguably seen in electron transport within large proteins [31], in bipartite protein-ligand interaction networks [19], in the mutualistic and trophic interactions of ecosystems [19] , and in human organizations and military field vehicle fleets [5].

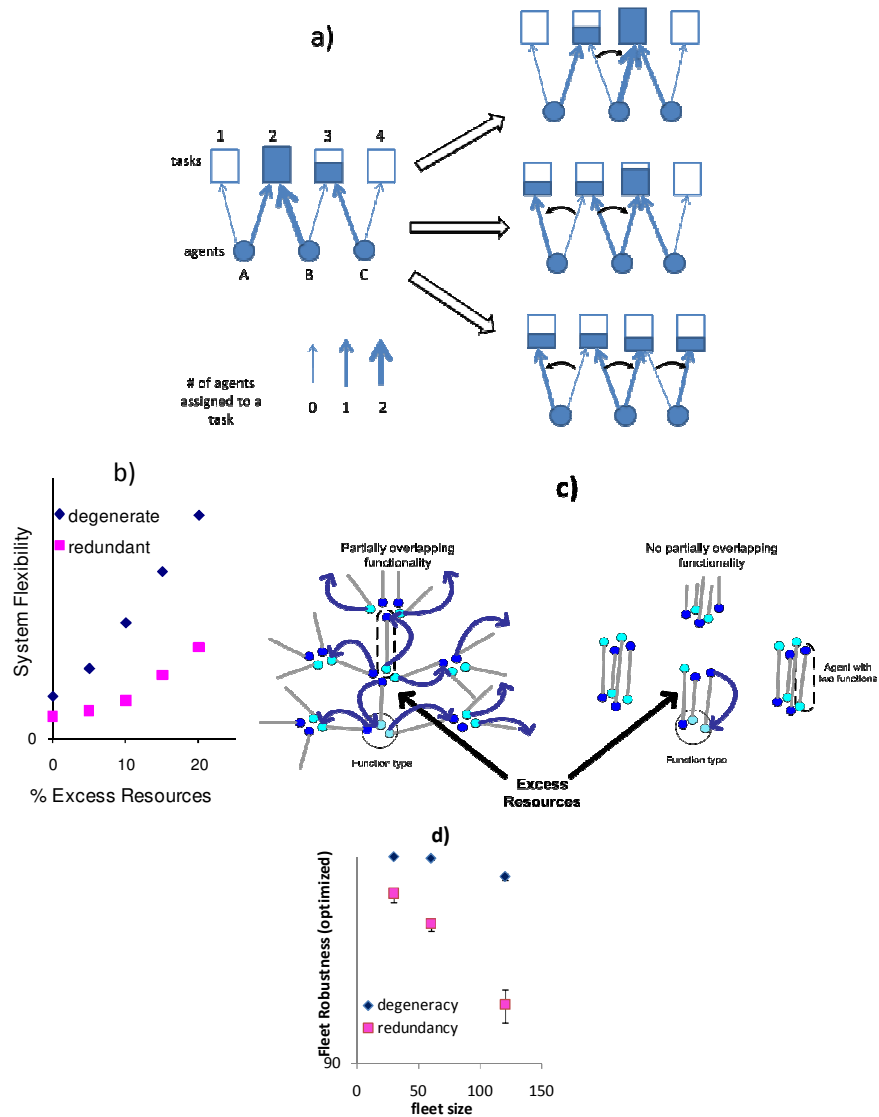

Figure 5 Panel a: The presence of unutilized type B agents can directly buffer fluctuations for task types 2 and 3. Type B agents are partly interoperable with type A and C agents, and thereby can free up resources for changes in task type 1 and 4 requirements.  Panel b: Flexibility conferred in random (non-optimized) protein affinity networks with increases in protein expression (excess resources) for protein affinity networks with and without degeneracy. Taken from [30]. Panel c: bi-functional agents designed to be partially interoperable (degenerate) with high flexibility conferred through a degeneracy backbone (left diagram) or fully interoperable (redundant) agents that are operationally isolated within functional clusters (right diagram). Adapted from [1].  Panel d: Robustness in fleets of multi-functional vehicles that were optimized to be maximally flexible towards a variety of mission scenarios. Quantitative differences in scenario task requirements are scaled proportionally with fleet size.  Adapted from [5].

For instance, we explored simulated vehicle fleets where partial interoperability in vehicle task capabilities was optimized to improve fleet responses toward anticipated variability in mission requirements. When partial interoperability was permitted to arise in the fleet design, we recorded improvements in fleet performance towards anticipated scenarios (Figure 5d) and greater robustness toward stresses that were not planned for during fleet development [5].  Interestingly adaptive responses that were optimized to address specific mission variations were also found to inadvertently support the emergence of pervasive flexibility options that could arise on demand and be co-opted on an ad-hoc basis to respond to conditions not previously encountered or planned for.

We showed in [19] that these new adaptive capabilities are likely due to the networked buffering effect just described (Figure 5c) and in [32] we proposed that similar principles can be applied to agile manufacturing processes involving semi-autonomous robots. Although this networked buffering effect increases the complexity of resource allocation decisions, we also found that the pervasive flexibility from NB endows a system with many high performance configurations. This desensitizes the system to sub-optimal local decisions thus partially negating conflicts between the system's adaptive response speed and the diversity contained in the response repertoire. In other words, NB partially resolves the robustness-efficiency conflicts and complexity-adaptation conflicts that plague engineered systems but are less prominent in biological networks.

One possible reason that this design approach is rarely tested is that multi-functional agents must share their time across several related functions and thus are not reliably available for each function they participate in (Figure 4c). However groups of partly interoperable agents counterintuitively display improved functional reliability at the group level as a result of resource allocation flexibility. This flexibility and associated reliability has been found to grow rapidly with gradual increases to excess resources (Figure 5b).

While the distributed flexibility we described relates to quantitative changes in a system's functional outputs, within a systems context this flexibility also becomes essential when exploiting exaptations or exploiting adaptations from response diversity. For instance, in efficient/optimized systems, selecting an adaptive option from a repertoire of diverse resources can reduce the availability of these co-opted resources for their original tasks thereby creating internal stresses on the what/when/where options for a system's other operational outputs. In such circumstances, task assignment flexibility becomes an important foundation for these higher level adaptive capabilities.

## 6. Recommendations for Enabling Degeneracy

*Shared Protocols, Agent Versatility, and Loose Coupling constitute a set of quantifiable design principles for realizing degeneracy and the emergence of pervasive flexibility in living technologies.*

Degeneracy and network buffering architectures can be incorporated into living technologies through the inclusion of clearly definable system features. These features have evolved in biological systems over long periods of time through major evolutionary transitions and have become ubiquitous in present day species, particularly in multi-cellular Eukaryotes, through repeated rounds of adaptive radiation [33] . In stark contrast to the fortuitous discovery of these properties within different biological contexts, we contend that these properties can be intentionally selected, designed, and encouraged. By enabling the systematic development of degeneracy, these properties support the development of living technologies that flexibly respond to unplanned changes at all scales of a system from operational environment and internal design to user preferences and competitive marketplace.

As a design principle, degeneracy can be realized in technological systems that exhibit the following features:

1. **Shared Protocols**
2. **Agent Versatility**
3. **Loose Regulatory Coupling**

We describe each of these design features while ignoring their biological and engineering motivations, history, and the technical jargon that pervades the research informing this discussion. An overview of these features is provided in Figure 6.

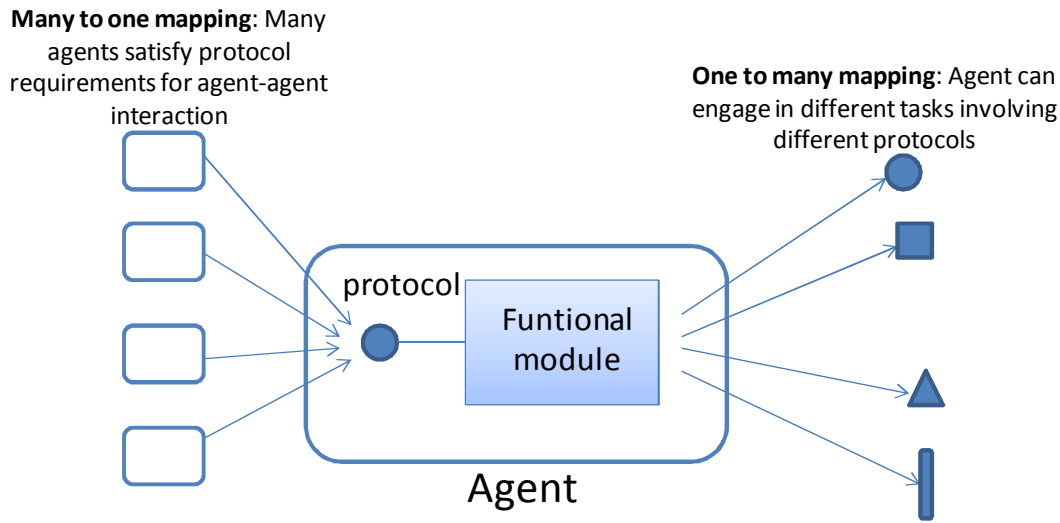

Figure 6: architecture of agent interactions and design that enable degeneracy. An agent consists of a protocol constrained input that determines activation/engagement by many structurally distinct agents, and a versatile functionality that allows the agent to engage in different functions. These "many agents-to-one function" and "one agent-to-many function" mappings are preconditions for realizing degeneracy and systemic flexibility through networked buffering (Fig 3).

## Shared Protocols

Plug-and-play compatibility provides unbounded opportunities for communication/interaction amongst technological artefacts. This supports the fortuitous discovery of novel service combinations and the occasional reorganization of networked services to reveal novel capabilities.

Such combinatorial flexibility is achieved in part by requiring agents to adhere to protocols. Protocols are standard procedures or "rules of engagement" [34] that specify conditions that must be met in order to execute a particular task or elicit a particular response/behavior in other agents [15, 34, 35]. Protocols enable non-trivial interactions amongst agents with little knowledge about the internal operations of the other. Instead, small amounts of information sharing between agents can be used to guide/inform elaborate patterns of action by each agent.

In agent-agent interactions, protocols manage relationships by defining precise rules for how the behavior of an agent is influenced by others. Although protocols constrain how agents can interact and communicate, they also desconstrain the potential for new collaborations because any agent can plug-and-play once it meets the protocol criteria. Within the context of the task-centric conceptual framework favoured in this article, protocols can be viewed as an agreed standard for the successful completion of tasks. From this perspective, protocols provide a design principle by which diverse components engage in functionally similar activities (functional redundancy) and thus help to manage a key requirement for the realization of degeneracy (**Error! Reference source not found.**a).

There are many examples of protocols that arise in different systems contexts. In biology, protocols have evolved and are seen in the energy currencies of metabolism, in cell-cell communication of neurons by activation potential, and in the universal usage of nucleotide sequence codons in gene transcription. In informal social systems, protocols emerge through shared acceptance of cultural norms that spread like viruses over socially connected and susceptible segments of a society. In technological systems, protocols are often explicitly established during system design, e.g. the internet's TCP/IP protocol stack.

The role of protocols in agent-based collaborations is not restricted to direct interactions. For instance, the manipulation of shared environmental artefacts using standards of manipulation (known to biologists as stigmergy) can provide cues for actions that are taken by other agents in the environment.

### Agent Versatility

Versatility describes the ability to competently perform a variety of partly related tasks or functions. When agents are functionally versatile, functions are invoked based on the agent's current state and cues from its local surroundings. Within a systems context, functional versatility might allow an agent to collaborate with or support ("plug into") a variety of different agents that each require engagement using different protocols.

The behaviour of versatile components is influenced by environmental cues and therefore can be responsive towards changing task requirements, e.g. changed requirements in operational outputs, replacement of degraded units, or replacement of units that have been assigned to other tasks. Pairs of versatile agents may appear functionally interoperable for certain tasks but uniquely functionally qualified for others, thereby enabling defining attributes of degeneracy to arise (**Error! Reference source not found.**). Conversely, degeneracy cannot be observed without functional versatility.

### Loose Regulatory Coupling

The ability to establish degeneracy within a system architecture is supported by the presence of shared protocols and loose regulatory coupling. Loose regulatory coupling refers to circumstances where the design features that determine agent behavioural responses (input protocols) are encapsulated and independent from design features that influence an agent's functional capabilities; see Figure 6. With loose regulatory coupling, design changes to agent functions will rarely require new input specifications or changes to the external cues that motivate agent activity. On the other hand, without loose regulatory coupling, agent design changes can alter an agent's protocols for engagement. This may in turn require collaborating agents to modify their behaviour, or worse may require changes in the design of collaborating agents that then propagate new requirements to still other agents of the system.

Loose regulatory coupling is a key innovation in gene regulation and some signalling pathways [36] and has played a major role in the evolvability of biological systems [33]. The lack of loose regulatory coupling increases fragility within interdependent systems towards even small changes in design because of propagating change requirements in other interacting components. Conversely, loose regulatory coupling enables co-option of degenerate elements for innovative use in new applications without the propagation of new design requirements, i.e. it is a vital facilitator of design exaptation by degeneracy.

## 7. Discussion

### Understanding Complexity in Living Technologies

Complexity is a poorly understood concept in science, in part because it has been attributed multiple imprecise meanings. While the term *complexity* generally relates to the interdependence of component behaviour/actions/functions, it is an otherwise ambiguous term and there is no consensus as to its meaning or measurement, e.g. [37] [38] [39] [40] [41] [42] [43]. In [44] they point out at least three different meanings commonly associated with complexity:

- Complexity of size: a large number of diverse components
- Complexity of interconnection: a large number of highly specific forms of engagement amongst components
- Complexity of interaction: system components are functionally versatile and interact with numerous other components in a manner that is tailored to the interaction context.

In engineering, complexity often refers to sophisticated services that require interdependent actions of single-purpose devices; each occurring in specific ways, places, and times. In other words, engineering complexity often relates to complexity of interconnection. In the absence redundancy and diversity, interconnection complexity can reduce a system's adaptive potential in a manner that is easy to appreciate.

Starting with a single device, the number and exactness of operational constraints/specifications will restrict the proportion of operating conditions that will meet these requirements. Although the trade-off between operating constraints and operational feasibility is not necessarily linear or monotonic, the reliability of many multi-device services become more fragile to novel internal and external conditions as more components are added that each co-specify the feasible operating conditions of others contributing to the service. In other words, the operating requirements placed on each device become more exacting as its function becomes more reliant on the actions/states/behaviours of others, e.g. through direct interaction, through sharing or modifying the same local resources, or indirectly through failure propagation. Services with this interconnection complexity can become more fragile to atypical component behaviours and atypical events because a greater proportion of events will exceed the operational tolerance thresholds in at least one device, with the propagation characteristics of these threshold-crossing events determining the likelihood of sub-system and system-wide failure. To reduce the frequency of failures, a design approach is sometimes taken that assumes predictability and relies on carefully placed backup devices, the monitoring of intermediate system states, and the incorporation of a variety of fail-safe procedures. Failures can still be prevalent in such systems however as discussed in detail for Xerox photocopiers [45] and for some large organizations that have experienced rapid change such as DuPont [46]. Design principles such as modularity and loose coupling can help reduce the size and frequency of some failures, however adaptation processes containing repertoires of system response options are still essential for achieving reliable performance under unexpected conditions.

The same conditions that limit operational robustness toward unanticipated events can also place limits on the adaptability of system design. When systems are designed from single purpose devices that are each uniquely suitable for a system-critical function, this establishes a tight coupling between system performance, the reliability of a function, the continued normal operation of the device providing that function, and the continued compatibility of that device with other interacting devices.[1] Novel redesign of devices is thus constrained by a need to properly interact/communicate with other specific devices. With engineering driven to maximize efficiency and performance, small design adjustments are repeatedly made over time to improve efficiency under standard operating conditions, i.e. the system's design becomes well-adapted to a specific and well-controlled environment. As a system's design matures (evolves) in this way, there can become fewer alternative system configurations for achieving each given task and fewer degrees of freedom for modifying a system's design without compromising function.

If the environment or system priorities were to substantially change, this creates a need for system redesign and a lack of useful redesign options can create tension that grows over time. Eventually a failure to meet system-level goals can make reengineering unavoidable, however with the redesign constraints just discussed, design modifications become necessary in many components simultaneously; a phenomena that is ubiquitous in complex engineering artefacts and nicely illustrated by the dramatic redesign of complex software, e.g operating systems. A large reengineering effort often runs roughshod over the accumulated and highly contextual knowledge that was built during its earlier maturation, causing many large reengineering and change management projects to appear as failures when compared to prior system performance.

## Complexity in Biology

Highly sophisticated services also exist in biological systems that require many different sub-functions and process pathways to be executed. However, the building blocks of biological systems are not single purpose devices with predefined functionality and instead display considerable overlap in function, functional versatility, and degeneracy.

While occasional slowdowns in the tempo of adaptation is inevitable and occurs in biological evolution as well (e.g. under stabilizing selection), there is little evidence to suggest that biological systems experience the same built-up tension from gradual changes in the environment or the same sensitivity to incremental design changes. We believe this is because degeneracy affords a weaker coupling between the functions performed and the components involved in achieving them [33]. Within an abstract design space or fitness landscape, one might say that engineered systems find themselves on isolated adaptive peaks where large movements in design space are needed in order to find new feasible/viable design options while biological systems reside on highly connected neutral plateaus. Although many complexity science researchers have used the rugged fitness landscape metaphor to advocate the need for disruptive and explorative search in the evolution of technologies, this is neither required nor observed in biological evolution.

In biological evolution, continued species survival requires that incremental adaptive design changes can be discovered that do not lead to a propagation of redesign requirements in other components in the system, i.e. macro-mutation is a negligible contributor to the evolution of complex species.

---

[1] While functional redundancy is sometimes designed into a system, it is almost always treated as a backup device that is not utilized under standard operating procedures.

Instead, single heritable (design) changes are found that lead to (possibly context-specific) novel interaction opportunities for a component, flexible reorganization of component interactions (that still maintain core functionalities), and in some cases a subsequent compounding of novel opportunities within the system [47]. In other words, the requirement is one of incremental changes in design and compartmentalized, but not necessarily incremental, changes in system behaviour. In their paper "Evolvability", Kirchner and Gerhart present a number of illuminating biological examples where this flexible reorganization takes place at cellular and developmental levels [33]. Degeneracy, and its associated complexity of interaction (multi-functionality) can support this flexibility through redundancy and diversity and sometimes by providing forms of distributed flexibility that can only be realized within a systems context. In short, based on the arguments outlined here, we propose that degeneracy could play a major role in the realization of pervasive flexibility within living technologies.

## 8. Swarm Robotics

Many of the aspirational properties of living technologies are also implicit goals in swarm robotics research. Swarms of simple cooperative robots are being studied for their ability to solve complex real-world tasks in which adaptive responses to unexpected conditions and component failure is essential to success. The swarm concept is inspired in part by social insects where local information, limited communication, and decentralized control can provide powerful emergent properties and robust problem solving skills.

Many studies have only considered robots that are identical in form and function and these swarms have been found to produce a limited repertoire of system behaviors. An important exception is the swarmanoid project [48]. The swarmanoid project has constructed systems of heterogeneous and dynamically connected autonomous robots. Through functional collaboration, communication, and assembly, the swarmanoid achieves multi-robot functions that are qualitatively distinct from the capabilities of the individual robots acting in isolation. Three types of robots have been created – eye-bots, hand-bots, and foot-bots – that cooperate in a swarm of dozens of robots to achieve goals involving multiple complex sub-tasks. Each robot in swarmanoid is functionally versatile and interacts with its environment in several non-trivial ways. For instance, the eye-bots are specialized for sensing and analyzing the environment but can also fly and magnetically attach themselves to a ceiling, thereby expanding the robot's sensing and communication capabilities. Hand-bots are able to use their hands in a versatile way; climbing, grabbing, and manipulating other robots. Foot-bots are specialized to move over rough terrain and can transport a variety of objects including other robots. The swarminoid system provides a proof of principle that swarms of functionally unique robots can cooperate to achieve complex tasks. Protocols for robot-robot engagement enable a plug and play architecture that, in principle, can be extended to integrate new robot designs into the swarm collective. It is also conceivable for novel functions to be discovered through new swarm configurations that are guided by new patterns of environmental cues. In short, important forms of operational adaptation and design evolution are attainable in swarmanoid.

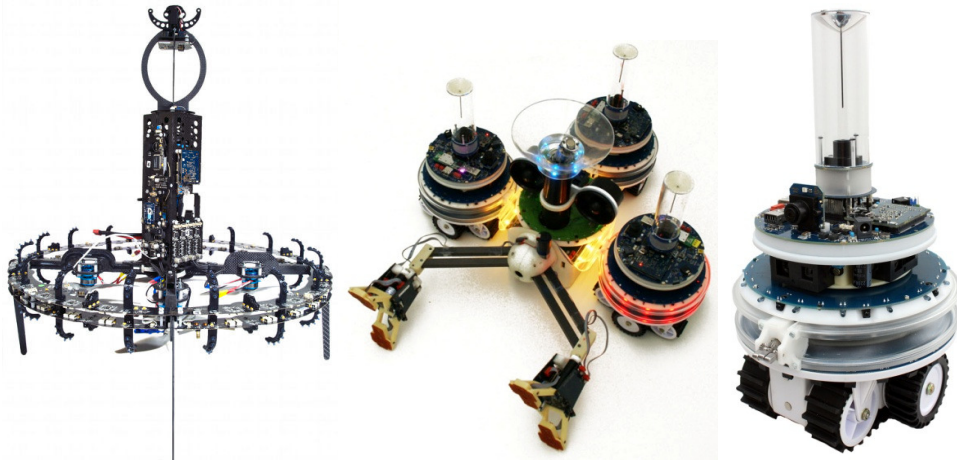

Figure 7 Photos of swarmanoid eye-bot (left), hand-bot being carried by three foot-bots (center) and foot-bot (right).

A video of the swarmanoid system in action won the 2011 AAAI video competition for exciting advances in artificial intelligence (http://iridia.ulb.ac.be/swarmanoid-the-movie). While constituting an exceptional advancement over existing swarm systems, the video would seem to suggest that the swarm relies greatly on the presence of well controlled environmental conditions in order to perform its tasks. Based on the principles outlined in this article, we propose a few changes to the swarmanoid system that could dramatically improve system flexibility and thereby improve swarm performance in a heterogeneous and uncertain environment.

Decomposing the swarmanoid system into its most simple functions (e.g. sensing, moving, climbing, flying), it is clear that the components that make up the robots exhibit high redundancy but relatively little degeneracy. The design of each functional component used in the construction of a swarmanoid robot is a result of difficult decisions involving tradeoffs between factors such as energy efficiency, dimensional constraints, functional range, strength, processor power, and durability. Selecting a design for each component of a robot represents a choice amongst design solutions on a multi-objective Pareto front where improvements in one objective are likely to have a negative impact on at least one other objective. Each design option for a component will correspond with a range of conditions in which the component could perform a given function (e.g. see Figure 3). Comparisons between component design options residing on the Pareto front may in some cases reveal a partial overlap in the conditions in which two designs are interchangeable, i.e. degeneracy. Similar functional overlaps can also arise in comparisons between entire robots and multi-robot assemblies.

To realize benefits from degeneracy, multiple distinct components should be added that are functionally redundant under conditions where a function is most commonly needed, while providing unique functional competencies in less common but still important conditions. Unlike the wastefulness of simple redundancy, the components should only remain in the system if they are able to regularly contribute to tasks: no components should be allowed to sit idle and be retained for a low chance contingency. For the flexibility afforded by degeneracy to actually be useful, it would also be necessary for the swarm to operate in a complex environment where many partially related tasks are required of the system. However once the swarm was forced to operate under these more realistic conditions, the value from degeneracy would become more apparent.

With a plug and play architecture introduced at the robot assembly level (e.g. plug and play of sensors, flying component, magnetic attachment, grabbing device, chassis, battery, etc), and with alternative designs provided for each component, large combinations of new robot configurations would be available to expand the structural and functional diversity of the robot swarm. This would result in a system that could no longer be easily decomposed into distinct classes of foot, hand, and eye-bots. Assuming that protocols for collaboration and robot-robot assembly were maintained, and assuming that robots could explore their functional limitations in a safe environment, this diversity should allow the swarmanoid system to complete complex tasks in a greater variety of operational environments. Robustness is a key determinant of the commercial viability of robot swarms and the added complexity from degeneracy would need to be gradually developed to ensure reliability in a variable environment. However through the incremental testing and evolution of new robot designs, we believe that degeneracy can contribute to the flexibility of the swarm and increase the swarm's commercialization potential for complex and dangerous environments such as rescue operations, mining, and space exploration.

Importantly however, degeneracy provides more than just functional reliability in a volatile environment. It also expands exaptation opportunities in which new multi-robot functions are discovered through their assembly and usage within novel environments. The new functions might not be optimal or even effective, however they provide useful information and guidance for the designer to create new component designs or new robot designs that can eventually expand system operations into an important niche environment and that could ultimately be important to the competitive success of the overall swarm. This narrative of environment-revealed adaptations followed by design modification is paralleled in biological evolution in the form of exaptations (co-option of existing components for new functions) followed by mutation-driven genetic accommodation/assimilation (fine-tuning of the new functions).

## 9. Conclusions

Our research into distributed robustness suggests that classic reductionist valuations of resource redundancy can sometimes be highly inaccurate and lead to opposite conclusions regarding the value of versatile assets that are deployed within a volatile and uncertain environment [5]. We claim that a better understanding of distributed robustness requires that we move beyond heuristics where robustness from diversity is attributed entirely to bet-hedging and portfolio-theoretic arguments. What we sorely lack is a principled approach to systematically designing living technologies with forms of robustness and flexibility that can emerge on demand [5].

The networked buffering described in this article illustrates one form of emergent robustness that can be understood and designed without knowing precisely where flexibility will be needed or what perturbations will be experienced. To achieve this networked buffering effect, elements in the system must display a partial overlap in functional capabilities: in some contexts providing functional redundancy while in others providing response diversity. In biology this unique type of group behaviour is known as degeneracy. In this article we have described how component-level functional versatility and network-level functional redundancy enable degenerate elements to facilitate exaptations. We have also described how the inclusion of protocols and loose coupling can enable the incremental evolution of highly degenerate systems. Our previous research on the influence of degeneracy in multi-agent systems suggests that it could broadly contribute to the flexibility of living technologies. In particular, we have found:

1. Degeneracy can lead to an emergent form of *systemic flexibility* through what we call Networked Buffering (NB). Our NB hypothesis [2] extends the value of diversity beyond that of portfolio theory towards a system-level relationship between diversity, performance, and resilience.
2. When organized into buffering networks, degeneracy can enhance a system's capacity to take advantage of opportunities originating from component changes; exploit novel environmental conditions; and mitigate the effects from unanticipated stresses. Stated differently, systems with inbuilt degeneracy can develop an innate capacity to deal with unforeseen, new challenges.
3. Degenerate systems can sometimes violate robustness-efficiency tradeoff constraints that are widely perceived as hard tradeoffs, e.g. physical conservation laws. The robustness-efficiency balance attainable in systems with innate degeneracy significantly outperforms that achievable in systems where degeneracy is absent.
4. The strategic advantage of highly degenerate multi-agent systems does not appear to degrade operational effectiveness. In particular, the benefits of NB appear to be attainable at very small cost in terms of management overheads when decision making is distributed across the system.

Degeneracy is a system property that can be clearly articulated and defined for any system comprised of functionally versatile elements. Currently, there are several research programs that are exploring how the degeneracy concept can be translated into design principles for the realization of more flexible and resilient systems in different disciplines [4-6, 49]. For instance, in Defense capability studies, we have shown using simulations that fleets of land field vehicles with high degeneracy in task capabilities can improve operational robustness within anticipated mission scenarios yet at the strategic level provides exceptional design and organizational adaptability for responding to unanticipated challenges [5, 50]. We are also looking at how the degeneracy concept can be translated in the design of more flexible manufacturing and assembly systems [6], and for better performance in population-based dynamic optimization [4]. Still others are using these concepts to understand some of the weaknesses of contemporary peer review processes [51] and the requisite conditions for embodied [52] and simulated artificial life [3, 53, 54]. Simulations of protein-protein interaction networks and simple genome:proteome mappings have also suggested that degeneracy plays a fundamental role in facilitating positive relationships between mutational robustness and evolvability in biology [19, 30]. As a source of heritable variation and exaptation opportunities, degeneracy is believed to be an essential component of Darwinian evolution in natural populations, cf [5] [21, 30].

In this article we looked at how these ideas can assist in realizing some of the desirable features of living technologies. To ground these ideas we focused on swarm robotics as a case study and evaluated a particular system known as Swarmanoid. In many ways, the Swarmanoid project captures key design principles for the realization of behavioral adaptation, configurational flexibility, and design evolvability that are rare outside of natural systems. Swarmanoid exemplifies a system architecture with loose coupling in component design and a plug and play modularity that facilitates dynamic assembly guided by environmental context. The swarmanoid system supports self-assembly, distributed decision making, and component diversity: features that have been sorely missing in most robotics research.

However, to fully realize its potential as an applied living technology within environments that are volatile and uncertain, the swarmanoid system needs a greater degree of flexibility for responding to real-world conditions where the environment can vary from one task instance to the next. The flexibility afforded by degeneracy could provide advantages in the design and operation of robotic swarms, while the plug and play architecture already present through most of the Swarmanoid system should enable this diversity to arise from a relatively small set of core building blocks.

**General Recommendations**

Deriving benefits from degeneracy requires a top-down valuation of options under different scenarios (i.e. what aspects of diversity are translating into a competitive advantage) and a bottom-up assessment of opportunity for the deployment of existing assets under a variety of new conditions. On the other hand, introducing degeneracy may involve additional design, training, and management overhead costs. To justify such changes, decision makers should consider:

- an assessment of the changes in capabilities that expanded component functionality confers;
- what cost containment is achievable from reusable training modules for skill development or from reusable physical modules in construction;
- what are the expected returns on investment, e.g. does the system naturally lend itself to networked buffering effects where robustness can be increased significantly with negligible losses in efficiency.

These decisions should be rational to individual stakeholders with incomplete information and without factoring in largely intangible benefits such as the potential for adaptation under unanticipated conditions. Enhancing a system's adaptive capabilities involves difficult decisions. Trade-offs between the cost of redundancy and the need for flexibility require careful choices that reflect expectations on the size and nature of future volatility.

Critics of nature-inspired design often claim that biological systems display costly levels of component (e.g. protein) complexity and gratuitous amounts of inefficiency. Degeneracy does indeed require considerable component diversity which can come with design costs and management overhead costs. Biological systems solve this problem by creating degeneracy through the versatile re-use of a small number of molecular building blocks. Modular technological systems with a plug-and-play architecture should similarly be able to achieve high levels of degeneracy at relatively low costs.

# REFERENCES

[1]     J. M. Whitacre, "Degeneracy: a link between evolvability, robustness and complexity in biological systems," *Theoretical Biology and Medical Modelling,* vol. 7, 2010.

[2] M. A. Bedau, et al., "Living technology: Exploiting life's principles in technology," *Artificial Life,* vol. 16, pp. 89-97, 2010.

[3] E. Clark, et al., "Degeneracy Enriches Artificial Chemistry Binding Systems," *ECAL, Paris,* 2011.

[4] J. M. Whitacre, et al., "The role of degenerate robustness in the evolvability of multi-agent systems in dynamic environments," in *PPSN XI*, Krakow, Poland, 2010, pp. 284-293.

[5] J. M. Whitacre, et al., "Evolutionary Mechanics: new engineering principles for the emergence of flexibility in a dynamic and uncertain world (http://arxiv.org/pdf/1101.4103)," *Natural Computing,* (in press).

[6] R. Frei and J. M. Whitacre, "Degeneracy and Networked Buffering: principles for supporting emergent evolvability in agile manufacturing systems," *Journal of Natural Computing - Special Issue on Emergent Engineering,* (in press).

[7] M. Kondoh, "Foraging adaptation and the relationship between food-web complexity and stability," *Science,* vol. 299, p. 1388, 2003.

[8] M. L. Siegal and A. Bergman, "Waddington's canalization revisited: Developmental stability and evolution," *Proceedings of the National Academy of Sciences, USA,* vol. 99, pp. 10528-10532, 2002.

[9] P. Csermely, "Strong links are important, but weak links stabilize them," *Trends in Biochemical Sciences,* vol. 29, pp. 331-334, 2004.

[10] P. Csermely, *Weak links: Stabilizers of complex systems from proteins to social networks*: Springer Verlag, 2006.

[11] A. Levchenko and P. Iglesias, "Models of eukaryotic gradient sensing: application to chemotaxis of amoebae and neutrophils," *Biophysical Journal,* vol. 82, pp. 50-63, 2002.

[12] N. Barkai and S. Leibler, "Robustness in simple biochemical networks," *Nature,* vol. 387, pp. 913-917, 1997.

[13] T. M. Yi, et al., "Robust perfect adaptation in bacterial chemotaxis through integral feedback control," *Proceedings of the National Academy of Sciences of the United States of America,* vol. 97, p. 4649, 2000.

[14] F. A. Chandra, et al., "Glycolytic oscillations and limits on robust efficiency," *Science,* vol. 333, p. 187, 2011.

[15] M. E. Csete and J. C. Doyle, "Reverse Engineering of Biological Complexity," *Science,* vol. 295, pp. 1664-1669, 2002.

[16] J. Stelling, et al., "Robustness of Cellular Functions," *Cell,* vol. 118, pp. 675-685, 2004.

[17] A. Wagner, "Robustness against mutations in genetic networks of yeast," *Nature Genetics,* vol. 24, pp. 355-362, 2000.

[18] A. Wagner, "Distributed robustness versus redundancy as causes of mutational robustness," *BioEssays,* vol. 27, pp. 176-188, 2005.

[19] J. M. Whitacre and A. Bender, "Networked buffering: a basic mechanism for distributed robustness in complex adaptive systems," *Theoretical Biology and Medical Modelling* vol. 7, 15 June 2010 2010.

[20] J. Macia and R. Solé, "Distributed robustness in cellular networks: insights from synthetic evolved circuits," *Royal Society Interface,* vol. 6, pp. 393-400, 2009

[21] G. M. Edelman and J. A. Gally, "Degeneracy and complexity in biological systems," *Proceedings of the National Academy of Sciences, USA,* vol. 98, pp. 13763-13768, 2001.

[22] M. Csete and J. Doyle, "Bow ties, metabolism and disease," *TRENDS in Biotechnology,* vol. 22, pp. 446-450, 2004.

[23] S. van Wageningen, et al., "Functional Overlap and Regulatory Links Shape Genetic Interactions between Signaling Pathways," *Cell,* vol. 143, pp. 991-1004, 2010.

[24] H. Kitano and K. Oda, "Robustness trade-offs and host-microbial symbiosis in the immune system," *Mol Syst Biol,* vol. 2, p. 2006 0022, 2006.

[25] D. N. Reznick and C. K. Ghalambor, "The population ecology of contemporary adaptations: what empirical studies reveal about the conditions that promote adaptive evolution," *Genetica,* vol. 112, pp. 183-198, 2001.

[26] F. Figge, "Bio-folio: applying portfolio theory to biodiversity," *Biodiversity and Conservation,* vol. 13, pp. 827-849, 2004.

[27] D. Tilman, "Biodiversity: population versus ecosystem stability," *Ecology,* vol. 77, pp. 350-363, 1996.

[28] D. E. Schindler*, et al.*, "Population diversity and the portfolio effect in an exploited species," *Nature,* vol. 465, pp. 609-612, 2010.

[29] S. J. Gould and E. S. Vrba, "Exaptation-a missing term in the science of form," *Paleobiology,* pp. 4-15, 1982.

[30] J. M. Whitacre and A. Bender, "Degeneracy: a design principle for achieving robustness and evolvability," *Journal of Theoretical Biology,* vol. 263, pp. 143-53, Mar 7 2010.

[31] A. Kurakin, "The self-organizing fractal theory as a universal discovery method: the phenomenon of life," *Theoretical Biology and Medical Modelling,* vol. 8, p. 4, 2011.

[32] R. Frei and J. Whitacre, "Degeneracy and networked buffering: principles for supporting emergent evolvability in agile manufacturing systems," *Natural Computing,* pp. 1-14.

[33] M. Kirschner and J. Gerhart, "Evolvability," *Proceedings of the National Academy of Sciences, USA,* vol. 95, pp. 8420-8427, 1998.

[34] J. Doyle and M. Csete, "Rules of engagement," *Nature,* vol. 446, pp. 860-860, 2007.

[35] J. C. Doyle*, et al.*, "The "robust yet fragile" nature of the Internet," *Proceedings of the National Academy of Sciences of the United States of America,* vol. 102, p. 14497, 2005.

[36] R. P. Bhattacharyya*, et al.*, "Domains, motifs, and scaffolds: the role of modular interactions in the evolution and wiring of cell signaling circuits," *Annual Review of Biochemistry,* vol. 75, pp. 655-680, 2006.

[37] M. Gell-Mann, "What is complexity," *Complexity,* vol. 1, pp. 1-9, 1995.

[38] P. Senge, "The Fifth Discipline: The Art and Practice of the Learning Organization," *Consulting Psychology Journal: Practice and Research,* vol. 45, pp. 31-32, 1993.

[39] D. W. McShea, "Perspective: Metazoan Complexity and Evolution: Is There a Trend?," *Evolution,* vol. 50, pp. 477-492, 1996.

[40] C. Adami, "Sequence complexity in Darwinian evolution," *Complexity,* vol. 8, pp. 49-57, 2002.

[41] J. P. Crutchfield and O. Görnerup, "Objects that make objects: the population dynamics of structural complexity," *Journal of The Royal Society Interface,* vol. 3, pp. 345-349, 2006.

[42] R. M. Hazen*, et al.*, "Functional information and the emergence of biocomplexity," *Proceedings of the National Academy of Sciences,* vol. 104, pp. 8574-8581, 2007.

[43] B. Edmonds, "Complexity and scientific modelling," *Foundations of Science,* vol. 5, pp. 379-390, 2000.

[44] D. L. Alderson and J. C. Doyle, "Contrasting views of complexity and their implications for network-centric infrastructures," *Systems, Man and Cybernetics, Part A: Systems and Humans, IEEE Transactions on,* vol. 40, pp. 839-852, 2010.

[45] P. S. Adler and B. Borys, "Two types of bureaucracy: Enabling and coercive," *Administrative Science Quarterly,* pp. 61-89, 1996.

[46] A. D. Chandler, *Strategy and structure*. Cambridge, MA: MIT press, 1993.

[47] A. Kurakin, "Scale-free flow of life: on the biology, economics, and physics of the cell," *Theoretical Biology and Medical Modelling,* vol. 6, 2009.

[48] M. Dorigo*, et al.*, "Swarmanoid: a novel concept for the study of heterogeneous robotic swarms," Technical Report TR/IRIDIA/2011-014, IRIDIA, Université Libre de Bruxelles, Brussels, Belgium2011.

[49] M. Randles*, et al.*, "Distributed redundancy and robustness in complex systems," *Journal of Computer and System Sciences,* vol. 77, pp. 293-304, 2010.

[50] A. Bender and J. M. Whitacre, "Emergent Flexibility as a Strategy for Addressing Long-Term Planning Uncertainty," in *NATO Risk-Based Planning Conference* Salisbury, UK, 2011.

[51] S. Lehky, *Peer Evaluation and Selection Systems: Adaptation and Maladaptation of Individuals and Groups through Peer Review*: BioBitField Press, 2011.

[52] J. A. Fernandez-Leon, "Behavioural robustness and the distributed mechanisms hypothesis," DPhil PhD Thesis, University of Sussex, 2011.

[53] S. Kerkstra and I. R. Scha, "Evolution and the Genotype-Phenotype map," Masters Thesis, University of Amsterdam, 2008.

[54] M. Mendao*, et al.*, "The Immune System in Pieces: Computational Lessons from Degeneracy in the Immune System," *Foundations of Computational Intelligence, 2007. FOCI 2007. IEEE Symposium on,* pp. 394-400, 2007.